\documentclass{article}

\usepackage[numbers]{natbib}

\usepackage[preprint]{nips_2018}

\usepackage[utf8]{inputenc} 
\usepackage[T1]{fontenc}    
\usepackage{hyperref}       
\usepackage{url}            
\usepackage{booktabs}       
\usepackage{amsfonts}       
\usepackage{nicefrac}       
\usepackage{microtype}      
\usepackage{graphicx}
\usepackage{amssymb}

\title{From Audio to Symbolic Encoding}

\author{
Shenli Yuan \\
Department of Mechanical Engineering,\\
Center for Computer Research in Music and Acoustics (CCRMA)\\
Stanford University \\
\texttt{shenliy@stanford.edu} 
\And
Lingjie Kong \\
Department of Computer Science \\
Stanford University\\
\texttt{ljkong@stanford.edu} 
\AND
Jiushuang Guo \\
Department of Statistics\\
Stanford University\\
\texttt{jguo18@stanford.edu} 
}

\begin{document}

\maketitle

\begin{abstract}
Automatic music transcription (AMT) aims to convert raw audio to symbolic music representation. As a fundamental problem of music information retrieval (MIR), AMT is considered a difficult task even for trained human experts due to overlap of multiple harmonics in the acoustic signal. On the other hand, speech recognition, as one of the most popular tasks in natural language processing, aims to translate human spoken language to texts. Based on the similar nature of AMT and speech recognition (as they both deal with tasks of translating audio signal to symbolic encoding), this paper investigated whether a generic neural network architecture could possibly work on both tasks. In this paper, we introduced our new neural network architecture built on top of the current state-of-the-art Onsets and Frames \cite{hawthorne2017onsets}, and compared the performances of its multiple variations on AMT task. We also tested our architecture with the task of speech recognition. For AMT, our models were able to produce better results compared to the model trained using the state-of-art architecture; however, although similar architecture was able to be trained on the speech recognition task, it did not generate very ideal result compared to other task-specific models. 

\end{abstract}

\section{Introduction}
Speech recognition and automatic music transcription have both been popular but challenging tasks in their fields, namely, natural language processing (NLP) and music information retrieval (MIR). There are various reasons why these two tasks have been challenging, some of them are common problems for both tasks, such as sensitive to the recording conditions, limited availability of the datasets, requirements of large memory for audio data, etc. There are also problems specific to one of the tasks. For example: speech recognition often deals with different accents, polyphonic music deals with the complex overlap of multiple harmonics in the acoustic signal from the concurrently active notes. Nevertheless, the two tasks share the very similar nature of translating certain audio signals to specific types of symbolic encoding: Speech recognition translate human spoken languages to word or phonetic transcription, while AMT transcribes music to symbolic music representations such as Musical Instrument Digital Interface (MIDI) \cite{moog1986midi}, an industry standard music technology protocol widely used for music encoding. 

\subsection{Architectures}
Inspired by the similarities between AMT and speech recognition, we developed a neural network architecture that could be trained to tackle both problems. For AMT task, this paper limits the scope to automatic piano transcription (APT), which means all the raw audios will be polyphonic piano recordings.

In this project, besides the baseline model, which consists of a convolutional stack and a bidirectional LSTM, we also implemented multiple variations of a neural network architecture built on top of the Onsets and Frames architecture. Self-attention, dilated convolutional stacks, highway network were introduced to the network. These variations, combined or individually, were able to outperform the original Onsets and Frames architecture during our test. In addition, we attempted a less successful modification of the architecture to use L2 loss for the purpose of achieving more accuracy in the time prediction. Two of the described architectures were used for the speech recognition task, although the results were not very ideal.  The detailed description of the architectures and result analysis would be covered in latter sections of the paper. 

\subsection{Terminologies}
Besides the terms mentioned above, there are other terminologies used in this paper that might cause confusion; they are listed below for reference:
\begin{itemize}

    \item onset (onset time): the time at which a note begins;
    \item offset (offset time): the time at which a note terminates;
    \item velocity: A measure of how much force / speed at which a note on a keyboard is played. A higher velocity corresponds to a louder sound; 
    \item mel-spectrogram: spectrum of mel scale frequencies of a signal as it varies with time;
    \item frame: the "time" in raw digital audio is always discretized, when converting the audio to time-frequency representation, we extract frequency information at the cost of lowering the time resolution, a "frame" describes the smallest unit in time domain after the conversion.
\end{itemize}

\section{Related Work}

Currently the state-of-the-art architecture for APT is the Onsets and Frames developed by \citet{hawthorne2017onsets}, which is a supervised neural network model transcribing piano recordings to MIDI. The proposed model, using deep convolutional and recurrent neural network (RNN), is designed to transcribe polyphonic music without prior information about the recording environment. There have been previous works using deep learning and neural networks for AMT \cite{kelz2016potential, sigtia2016end}. They were both inspired by models used for other tasks. For example, convolutional neural network (CNN) for image classification, or a combined CNN and RNN model commonly used for speech recognition. However, Onsets and Frames has an unique design where the model emphasizes on the note onset detection. A dedicated note onset detector was trained and the output of the note onset detector is used as additional feature input for the frame wise note activation detector.  The reason behind such design is because the onset frame of a piano note is at the note’s peak amplitude, followed by relatively sharp decay and therefore easier to identify.

Deep learning approach for speech recognition was studied and widely used for natural language processing before it was used for AMT. Model like Connectionist Temporal Classification (CTC) uses a softmax layer to define a separate output distribution $Pr(k|t)$ at every step $t$ along the input sequence \cite{graves2013speech}. This distribution covers the number of possible phonemes plus an extra blank symbol $\varnothing$. Different from the onsets and frames architecture, which utilizes an onset network to only look for note present or absent and then an separated frame network to inference the note, CTC for human language allows the network to decide whether or not to emit any label, at every time step. Unlike piano notes for which the richest information resides in the onsets, speech signal has a relatively uniform distribution in its amplitude over time, which further complicates the problem.

CTC defines a distribution over phoneme sequence that only depends on the acoustic input sequence. Therefore, it is an acoustic-only model. A recent update, known as RNN transducer \cite{graves2012sequence}, combines both a CTC network with an independent RNN that predict each phoneme given the previous predicted one. This overall yields a language and acoustic model. Furthermore, with the advancement on enhancing RNN conditioned on input data through an attention mechanism in machine translation, image caption generation, and handwriting synthesis, attention has also been extended for speech recognition \cite{NIPS2015_5847}

\section{Approach}
This section summarizes the approaches taken to achieve our goals. For clarification, anything different from the original Onsets-and Frames architecture was implemented by our group, including the baseline network, the colored blocks in Figure~\ref{diagram_new} and Figure~\ref{diagram_l2}, modifications made for speech recognition tasks, data down sampling of MAESTRO dataset, data pre-process of TIMIT dataset, and PER evaluation metric for speech recognition. 

\subsection{Baseline}

\begin{figure}[ht]
\begin{center}
\includegraphics[width=2.5cm]{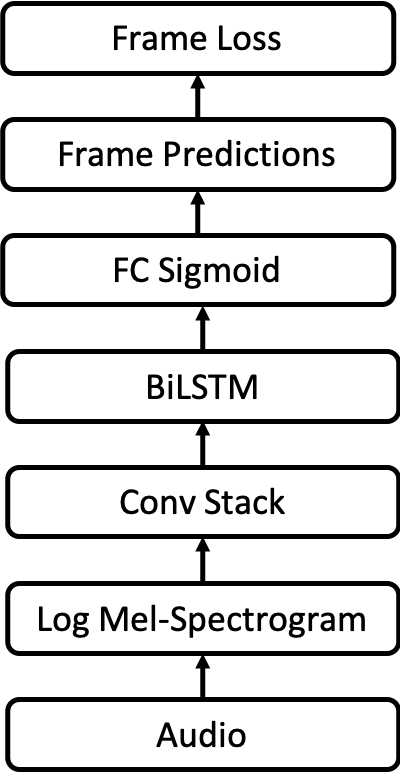}
\end{center}
\caption{Baseline network}
\label{diagram_baseline}
\end{figure}
The baseline model is based on CNN and Bidirectional LSTM as shown in figure~\ref{diagram_baseline}. This model first convert the raw audio (waveform) to a time-frequency representation (Log Mel-Spectrogram in this case), and use it as the input data of neural networks. Specifically, each audio data is divided into 20-second pieces with F = 625 frames per piece. In each frame, time serial data is converted into frequency representation using Discrete Fourier Transform (DFT) with bin size (number of frequency bins) B = 229. The input for Conv Stack is $ x \in R^{F\times B} $ while output for Conv Stack  $ x_{conv} \in R^{F\times K} $ in which K is a hyper-parameter we used to define our model complexity. For our model, K is picked to 256. Each cell of BiLSTM is $h_i = [\overleftarrow{h_i}, \overrightarrow{h_i}]$ where $ h_i \in R^{K\times 1} $ and output of BiLSTM is $x_{BiLSTM} \in R^{F\times K}$. FC layer output have $x_{FC} \in R^{F\times P}$ where P = 88 are total number of piano notes. Sigmoid function is used to cast output between 0 and 1 and cross entropy loss function is used to minimize the frame loss.

\subsection{Main approach}

\begin{figure}[ht]
\begin{center}
\includegraphics[width=13cm]{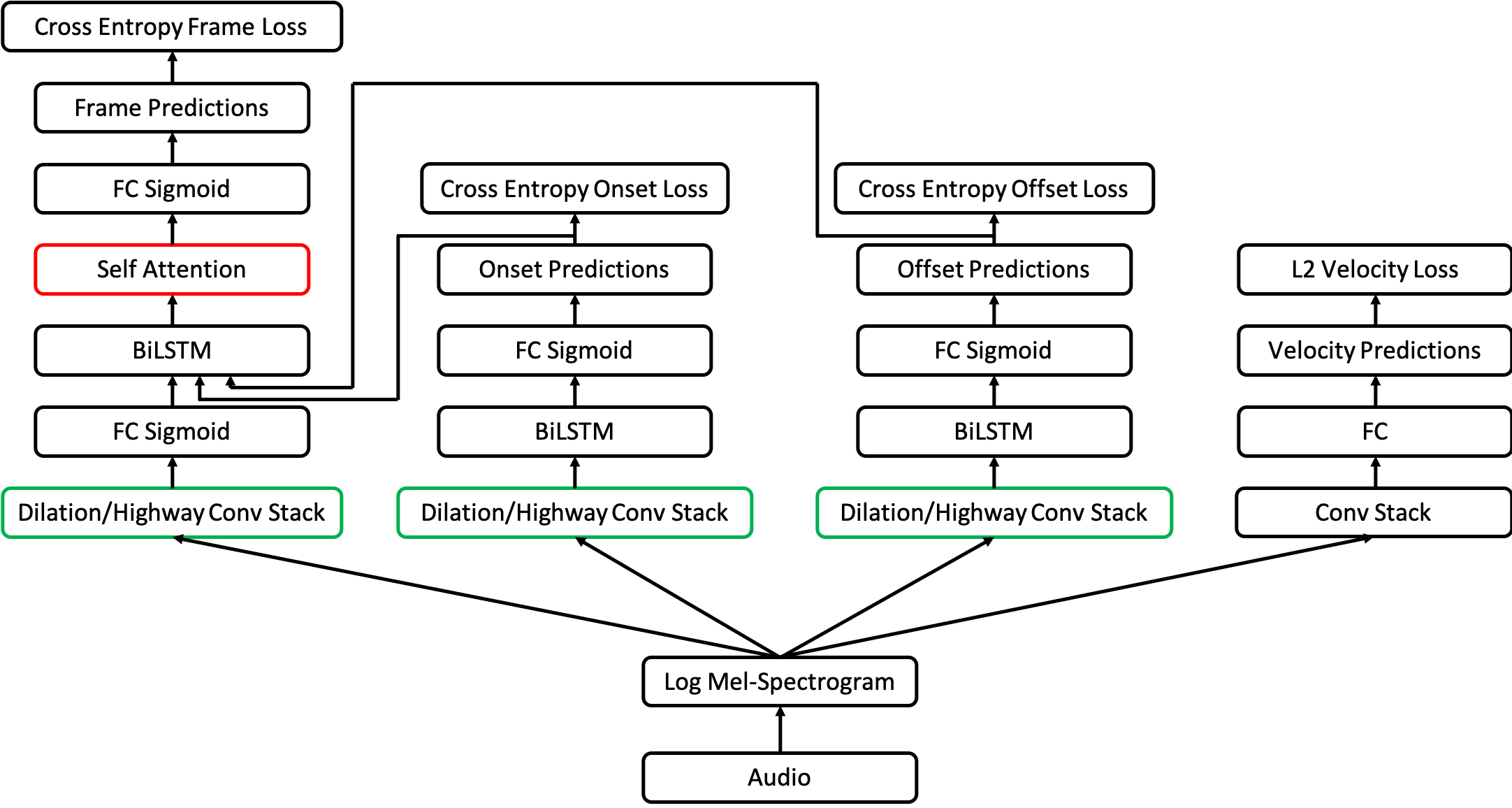}
\end{center}
\caption{Onsets and Frames + Dilation + Highway + Attention}
\label{diagram_new}
\end{figure}

Our model was built on top of the Onsets and Frames architecture (introduced in the Related Work section) with various additions/modifications as shown in Figure~\ref{diagram_new}. The variations were implemented individually and were combined afterwards for comparison. 

\subsubsection{Onsets and Frames with highway}
Network training becomes difficult as its depth increases. Highway Network \cite{srivastava2015highway} introduces a new architecture design to ease this problem of gradient-based training. It allows unimpeded information flow through layers and enables the network to only capture the difference. We add a convolutional layers with highway network in addition to each of the original convolutional stacks.

\subsubsection{Onsets and Frames with dilation}
\begin{figure}[ht]
\begin{center}
\includegraphics[width=10cm]{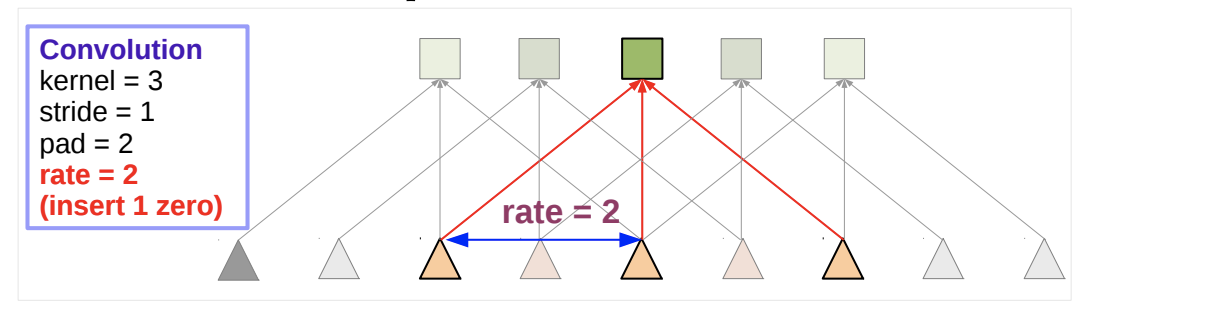}
\end{center}
\caption{dilation layer with dilation rate r = 2}
\label{diagram_dilation1}
\end{figure}

The classical Onsets and Frames model uses consecutive convolutional layers kernel of size 3, followed by a max-pooling and a ReLU layer in its convolutional stacks. This allows the model to learn high level abstract feature representations, but barely captures long distance relations between features, especially in computer vision related tasks due to the scaling or transformation of the images. In AMT or Speech recognition, this problem is still crucial, since the audio data was converted to 2-dimentional time-frequency representations (just like images). The time domain information will often be scaled due to various speed of the music or the speech.  In \cite{chen2018deeplab}, dilated convolutional layer is implemented to enlarge the field of view of filters. This enables the model to be more robust to scaling and capture long distance feature relations in a more efficient way. Dilation convolution works by skipping nearby features regularly at a constant rate of r (dilation rate).The standard convolution has dilation rate r = 1.  Formula of a 1-D input signal x[i] with a filter w[k] of length K is defined as $$ y[i] = \sum_{k=1}^{K} x [i + r \cdot k]w[k]. $$ Figure \ref{diagram_dilation1}\footnote{http://liangchiehchen.com/projects/DeepLab.html} is an example of a single dilation layer with r = 2 for 1-D data. Dilation convolution can increase the field-of-view (FOV) of filters in convolutional layers, which would efficiently  enlarge kernel size of a  $k\times k$ filter in normal convolutional layer to $k_{e}=k+(k-1)(r-1)$ without increasing number of parameters and computational complexity. 
In our experiments, we added the dilation architecture after regular convolutional layers in convolutional stacks. The result from regular convolutional layers is passed into three dilation layers with different dilation rate r (r = 1,4,8), the results from these three layers were averaged as the final result of the dilation layer.

\subsubsection{Onsets and Frames with Attention}
With the advancement on enhancing RNN conditioned on input data through an attention mechanism in machine translation, image caption generation, and handwriting synthesis, attention has also been extended for speech recognition \cite{chorowski2015attention}. Self-attention approach is also widely used in many tasks \cite{wang2018non} \cite{vaswani2017attention}. We believe that by using attention, our model could better capture the relationship of information among different frames. As for the architecture that we implemented, self-attention is added on top of Bidirectional LSTM and its output is concatenate with BiLSTM hidden layers as the input for next FC Sigmoid layer. 

\subsection{Onsets and Frames with different loss function}
\begin{figure}[ht]
\begin{center}
\includegraphics[width=13cm]{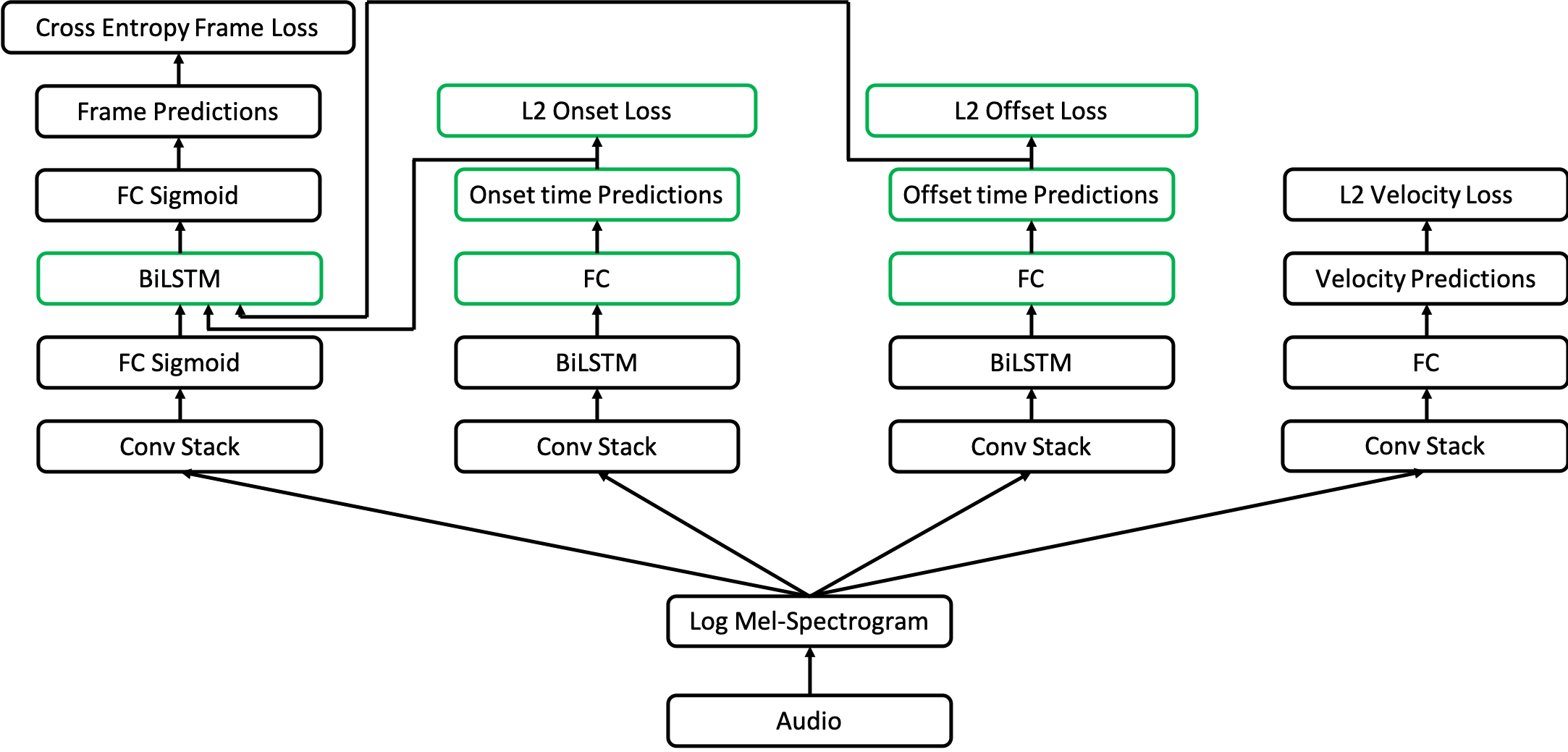}
\end{center}
\caption{Onsets and Frames + with different L2 loss}
\label{diagram_l2}
\end{figure}

The previously proposed architecture classifies whether certain note exists in certain frames, and the onset and offset times correspond to the frames at which a note starts and terminates. The architecture introduced in this section (Figure~\ref{diagram_l2}) tries to estimate onset and offset times of the notes directly through minimizing the loss between the ground truth and predicted onset and offset times, instead of trying to extract time information from the frame numbers. In this way, the time estimation would be less limited by the time resolution of the log mel-spectrogram; and each frame will have information regarding whether it has a note on, and its exact onset and offset times. 

\subsection{Architectures for speech recognition}
We used the onsets and frames architecture (without the velocity stack) as well as the baseline architecture to train the TIMIT dataset. The phonetic transcriptions were first translated into numbers and treated as midi notes for training and evaluation. The velocity stack of the onsets and frames architecture was removed because the velocity parameter does not exist in the phonetic transcriptions. After that, the baseline architecture was used instead of other more complex architectures that we implemented in this paper, because the phonetic information does not decay after its onset, and we believe that having the onset and offset stacks in the architecture would not improve the results. 

The onset and offset losses were removed from the calculation during training because the concepts of onset and offset are much weakened for speeches. 

\section{Experiment}
\subsection{Data}
The MAESTRO dataset \cite{maestro2018} was used for APT in this paper. It contains 1184 performances, approximately 430 compositions with 172.3 hours of total audio hours. The dataset was just released in 2018, thus has not been extensively used for similar tasks, but it is an order of magnitude larger than other commonly used dataset (e.g. the MAPS dataset). The dataset is divided into 80$\%$ of training data, 10$\%$ of validation data and 10$\%$ of test data. Due to the short time frame of the project, we sampled a quarter of the full dataset to be used for training, validation and test.

The TIMIT dataset \cite{garofolo1993timit} was used for speech recognition task in this paper. It contains 6300 phonetically rich sentences spoken by 630 speakers with 10 sentences for each speaker. Geographically, the 630 speakers are from 8 major dialect regions of the United States.The speeches are recorded as 16-bit .wav files with sampling rate of 16kHz. In addition to the audio files, the TIMIT corpus also includes time-aligned orthographic, phonetic and word transcriptions. Only the phonetic transcriptions out of the three were used in this paper; the phonetic transcriptions includes the start time, end time and the corresponding phoneme, which is very similar to the onset time, offset time and note in midi files. However, it does not contain velocity parameter in the transcription, unlike the midi transcriptions used in MAESTRO dataset. The TIMIT dataset was divided into training and test subset. We divided the test subset in half for test and validation, respectively. 

\subsection{Evaluation method}

For APT, frame-level and note-level precision, recall and F1 score were used as our evaluation metrics. This is done through $mir\_eval$\footnote{https://github.com/craffel/mir\_eval} python library for computing common heuristic accuracy scores for various music and audio information retrieval and signal processing tasks. In detail, precision and recall are calculated based on following criteria: 1) the predicted piano notes match the targets, 2) predicted onset and offset times are within 50ms of the target onset and offset times.

For speech recognition, phoneme error rate (PER) is used to evaluate our model. The reason why a different metric is used for TIMIT dataset is due to the irrelevance of onset and offset times in speech recognition. Phoneme error rates are calculated by using Levenshtein algorithm\footnote{https://github.com/jitsi/asr-wer/tree/master/jiwer}. It is defined as the edit distance between reference sequence and hypothesis divided by the number of elements in reference. 

\subsection{Experimental detail}
For the various APT models implemented in this paper, the raw audio was converted to mel-spectrogram with 229 frequency bins ranging from 20Hz to 8000Hz and a sample rate of 16000Hz. The model runs 500000 iterations with batch size of 4 and sequence length of 327680 for each batch. Learning rate of the model is 0.0006, decaying with rate 0.98 for every 10000 iterations. This model also implements L2 gradient clipping of each parameter at 3. The training time varies from \textasciitilde{7} to \textasciitilde{14} hours depending on the model complexities as well as early stop results.  

For the speech recognition, the converted mel-spectrogram have frequencies ranging from 30Hz to 300Hz due to a narrower frequency band of human speeches compared to piano sound. The rest of the parameters were the same as the ones used for APT.

\subsection{Results}

The results for the different architectures that we implemented are presented in Table~\ref{results-table}. The notations used in the tables are listed below:
\begin{itemize}
    \item note-w-o: Note with offsets
    \item note-w-v: Note with velocity
    \item note-w-ov: Note with offsets and velocity
    \item B+O:; baseline with the addition of the onset stack (no offset stack)
    \item OaF: the PyTorch implementation\footnote{https://github.com/jongwook/onsets-and-frames} of the original Onsets-and-Frames architecture 
    \item OaF+H: Onsets-and-Frames architecture with highway network
    \item OaF+D: Onsets-and-Frames architecture with dilated convolutional stacks
    \item OaF+H+D: Onsets-and-Frames architecture with both highway network and dilated convolutional stacks
    \item OaF+A: Onsets-and-Frames architecture with self-attention
\end{itemize}

\begin{table}
  \caption{F1 scores of different architecture for AMT task} 
  \label{results-table}
  \centering
  \begin{tabular}{llllll}
    \toprule
    \cmidrule(r){1-2}
    architecture & note      & note-w-o  & note-w-v  & note-w-ov & frame \\
    \midrule
         baseline & 0.698     & 0.523     & 0.683     & 0.514     & 0.514     \\
         B+O      & 0.890     & 0.688     & 0.864     & 0.673     & 0.856     \\
         OaF      & 0.898     & 0.708     & 0.872     & 0.693     & 0.870     \\
         OaF+H    & \bf 0.917 & \bf 0.736 & \bf 0.893 & 0.721     & \bf 0.886 \\
         OaF+D    & 0.904     & 0.714     & 0.879     & 0.699     & 0.865     \\
         OaF+H+D  & 0.915 & \bf 0.736 & 0.891     & \bf 0.722 & 0.884     \\
         OaF+A    & 0.911     & 0.732     & 0.878     & 0.714     & 0.881    \\
    \bottomrule
  \end{tabular}
\end{table}

\begin{table}
  \caption[Training time for different architecture for AMT task; (early stop criteria: max trial = 10, max patience = 10)]
    {\tabular[t]{@{}l@{}}Training time for different architecture for AMT task \\ (early stop criteria: max trial = 10, max patience = 10)\endtabular}

  \label{time-table}
  \centering
  \begin{tabular}{ll}
    \toprule
    \cmidrule(r){1-2}
    architecture  & training time (hh:mm:ss)      \\
    \midrule
         baseline & {03:25:39} \\
         B+O      & {05:11:35} \\
         OaF      & {06:09:03} \\
         OaF+H    & {09:26:43} \\
         OaF+D    & {11:13:17} \\
         OaF+H+D  & {11:52:34} \\
         OaF+A    & {04:41:23} \\
    \bottomrule
  \end{tabular}
\end{table}

\begin{figure}[ht]
\begin{center}
\includegraphics[width=10cm]{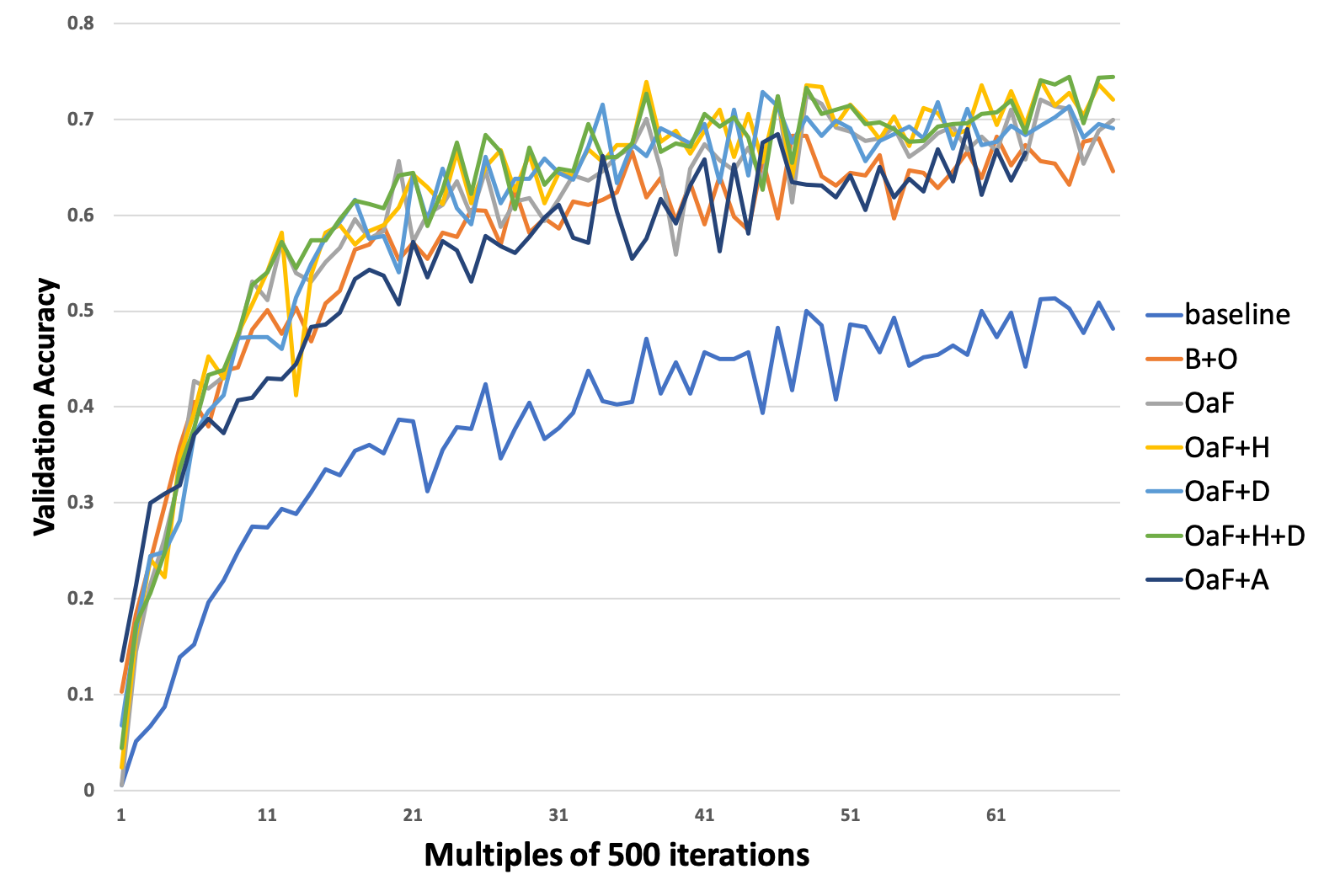}
\end{center}
\caption{Validation accuracy vs training iterations}
\label{validation_accuracy}
\end{figure}

The most direct observation from the result comparison is that the variations that include highway networks have the best performance. In addition, the onset stack serves a more important purpose as the onset stack, as the addition of onset stack to the baseline architecture greatly improves the results. The results from the architecture with L2 loss were not included in this table due to its poor performance. For the speech recognition, we used phoneme error rate (PER) as the evaluation metric instead of F1 scores. We were able to achieve \textasciitilde0.9 PER, which is much higher than lots of task-specific architectures. The results would be further analyzed in the next section. 

\section{Analysis}

From the results presented in the previous section, the highway network and the dilated convolutional stacks clearly improve the model performance so that our architecture was able to outperform the state-of-art architecture by a few percent. The highway network adds robustness against the depth of the neural networks; and the dilation increases the model's ability to capture longer distance feature relations, so that the model would be more robust against scaling (commonly happens in time domain for audio data). 

We had a relatively high hope for self-attention, and it did end up improving the model, but not as much as we expected. This could result from the difference between natural languages and music. In natural language, each word highly depends on the context it resides in, and attention helps each word capture relationship with others. The notes in music, on the other hand, are independent from each other. This means that a note at certain time step does not really affect which notes come before or after.

The architecture with L2 loss did not perform very well because the duration when a note is off is much longer than the duration when a note is on. When a note is off, both its onset and offset time will be labelled as "0". Therefore, compared to small amount of the onset and offset times of a note, prediction that a note is off in each frame is also included in the calculation of L2 loss function, which, due to its much larger amount, dominated the loss function and resulted in poor results. 

The speech recognition task was not quite successful due to various reasons. First, unlike music notes which have large magnitude when at start, speech phoneme is more uniformly distributed in time domain. This will affect the prediction accuracy of onset network, and subsequently the frame accuracy. Secondly, for music notes, although the magnitudes decay rapidly after onsets, their spectral patterns (frequency contents) remain similar. On the contrary, although the magnitudes of phonemes do not necessarily decay with time, their spectral patterns are constantly changing. This makes identifying the same phoneme over frames more difficult. Moreover, our architecture does not include the detection of duplicated phonetic prediction over frames, thus the duplicated predictions cannot be removed.

\section{Conclusion and future work}

In this project, we developed a neural network architecture that was able to outperform the state-of-art architecture for AMT task. We investigated multiple variations of the architecture and compared their performance. We were able to train models using the same architecture for the task of speech recognition, but the result were not ideal. 

For future development, post-process for speech recognition could be added to remove duplicated phonetic predictions. Connectionist Temporal Classification loss function should be used to minimize the loss of onsets and offsets times.  

Furthermore, for most audio to symbolic encoding, a fixed frame window size is used in DFT to convert time series data into frequency domain before feeding it into the neural network. Therefore, there is a trade off on how fine the window size should be to discretize the input audio. Specific domain knowledge is needed to determine what might be the best window size for music as well as human language. Therefore, neural networks can be added to learn how fine the window size should be instead of treating it as a hyperparameter. This on-the-fly DFT concept will slow down the network training because time serial data cannot be pre-processed. However, it will ease the problem of limited information in each frame due to fixed sized DFT windows.

\section{Acknowledgement}
We would like to thank the mentor of the project - Annie Hu, for the feedback that she provided for the project proposal and milestone, as well as the advice she offered through emails and during our meeting. We would also like to thank the entire CS224N teaching team for their dedication and guidance that made the project possible.  

{\small
\bibliographystyle{unsrtnat}
\bibliography{bibliography}
}


\end{document}